\documentstyle[prl,aps,amsfonts,amssymb,floats,epsfig,twocolumn]{revtex}
\begin{document}
\draft
\twocolumn[\hsize\textwidth\columnwidth\hsize\csname
@twocolumnfalse\endcsname
\title{Does EELS haunt your photoemission measurements?}
\author{K. Schulte, M. A. James, P. G. Steeneken, G. A. Sawatzky }
\address{Material Science Center, University
 of Groningen, Nijenborgh 4, 9747 AG Groningen, The Netherlands}
\author{R. Suryanarayanan, G. Dhalenne, A. Revcolevschi}
\address{Lab. de Chimie des Solides, B\^{a}t. 414, CNRS, UA 446,
Universit\'{e} Paris-Sud, 91405 Orsay, France}
\date{\today}
\maketitle
\begin{abstract}
It has been argued in a recent paper by R. Joynt (R. Joynt,
Science {\bf 284}, 777 (1999)) that in the case of poorly
conducting solids the photoemission spectrum close to the Fermi
Energy may be strongly influenced by extrinsic loss processes
similar to those occurring in High Resolution Electron Energy Loss
Spectroscopy (HR-EELS), thereby obscuring information concerning
the density of states or one electron Green's function sought for.
In this paper we present a number of arguments, both theoretical
and experimental, that demonstrate that energy loss processes
occurring once the electron is outside the solid, contribute only
weakly to the spectrum and can in most cases be either neglected
or treated as a weak structureless background.
\end{abstract}
%
\vskip2pc]
\narrowtext
\section{INTRODUCTION}
Photoemission has been used for years as a reliable technique for
probing the the electronic structure of the occupied states in
solids ranging from insulators through semiconductors and metals
to superconductors. In his article Joynt provided very convincing
and interesting arguments that especially for badly conducting
samples (roughly $\rho_0 \geq 0.1 m\Omega cm$, the Mott value) the
photoelectron spectrum may be affected by energy loss structures
resulting from the interaction with the time dependent fields set
up by the photoelectron receding from the surface of the solid. He
argued that the influence of these loss processes can be so strong
that the spectrum is dominated by them and that therefore the
intrinsic information regarding the electronic structure of the
solid all but disappears. Since photoemission is playing such a
prominent role in the discussion of strongly correlated materials
like the High Tc's, or more generally the transition metal oxides,
as well as in Kondo and Heavy Fermion systems, it is of quite some
importance to further investigate Joynt's assertions. In this
paper we study Joynt's arguments and provide both experimental and
theoretical findings that show that the effects due the losses
discussed by Joynt are only a small contribution to the total
spectrum and the zero energy loss probability for the
photoelectrons dominates for samples of either good or bad
conductivity.

\section{INTRINSIC AND PSEUDO-INTRINSIC EFFECTS}
Compared to other techniques, photoemission provides the most easy
and direct measurement of the one electron Green's function and
the directly related occupied density of states of a solid, if one
keeps in mind the possible influence on the photoelectron of a
number of `pseudo-intrinsic' effects\cite{Sawatzky}. By this we
mean first of all the matrix element needed for the description of
the amplitude of the optical transition probability from an
occupied state to a high energy unoccupied state in the solid,
followed by energy loss processes occurring on the electron's path
to the surface and finally the description of the escape of the
electron through the surface region into the vacuum. Where we
emphasize that in the so called `sudden approximation' we neglect
all interactions and interference effects between the high energy
excited electron and the hole left behind.\\
Also one has to contemplate the truly intrinsic effects: in a one
electron approximation the associated one hole Green's function is
a delta peak at an energy determined by the band dispersion of the
occupied states. In reality the electrons in the solid are usually
not simple free electrons but they interact with other electrons,
phonons, magnons {\it etc.}, resulting in one electron Green's
functions now including a frequency and momentum dependent
self-energy. For weakly interacting systems the initial delta
function spectrum for such an electron broadens (asymmetrically)
and attains a frequency distribution for each momentum vector,
which basically provides information not only of the
quasi-particle dispersion and lifetime, but also of the way the
electron is dressed inside the solid, due to the response of its
environment to its presence or absence. In strongly interacting
systems this self-energy causes a rather large spreading out of
the initial delta peaks describing the one hole Green's function,
and the description in terms of a quasi-particle with a certain
lifetime may break down completely. In these cases it is indeed
difficult to separate the intrinsic properties of the one hole
Green's function from the pseudo-intrinsic effects due to the
energy losses suffered by the excited electron on its way out of
the solid. These losses are basically dominated by the self-energy
of the excited electron. It is extremely important therefore to
have good estimates of the contributions due to energy loss
processes to the spectrum so that these can be identified, and
possibly corrected for, if they are substantial.\\
\\
Experimentally there are several ways of checking the expected
influence of these energy loss processes. The most obvious is to
study the energy loss spectrum of electrons incident on the solid
with an initial kinetic energy equal to that of the escaping
photoelectron. These loss spectra provide one with direct
information about the self-energy of an excited electron in the
unoccupied states of the solid. For high energy electrons
(E$_{k}\geq$ 60 keV) a transmission EELS measurement is possible
in which the obtained energy loss information is mostly due to
bulk processes. However, to achieve high energy resolution,
photoemission measurements are usually performed at low photon
energies (E$_{Ph} < 100$ eV) such as HeI radiation (21.22 eV) or
even less. At these low energies electron energy loss processes
can be studied with reflection High Resolution EELS but one must
then realize that the incident electron hardly penetrates the
solid surface and that therefore the surface loss function is
measured rather than the bulk losses experienced by a
photoelectron originating from inside the solid as in
photoemission. The surface and bulk loss function are linked to
each other however and thus information about the intrinsic
processes can be retrieved also from a reflection experiment,
although the relative intensities of certain bulk losses compared
to their surface analogs may differ. We note here that Joynt
concentrates on losses occurring after the photoelectron has left
the solid and so these should be directly related to the
reflection ELS losses.\\
\\
Another way of getting the self-energy information of the excited
electron is to study the spectral function in a photoemission
measurement of a narrow atomic core level of the solid at a photon
energy such that the excited electron will have the required (low)
kinetic energy. Although this loss spectrum will be intertwined
with the satellite structure due to the self-energy effects
involved in the sudden creation of the core hole itself, they will
nonetheless provide us with an upper limit on the importance of
the materials energy loss contributions to a photoemission
spectrum. This was in fact suggested and used in
Ref.~\cite{Sawatzky} to argue that the broad and intense energy
distribution seen in angular resolved photoelectron spectroscopy
of the high Tc's was not a result of energy loss processes but
mainly  a direct result of the strongly energy dependent
self-energies in these strongly correlated materials. A more
detailed study of this effect in the High Tc's has recently been
published\cite{Norman}.\\
\\ Thus, in the interpretation of spectra, it is generally assumed
that the photoelectron intensity distribution is a true reflection
of the single particle spectral function, and that the
aforementioned pseudo-intrinsic losses can be identified and
reckoned with if necessary.
\section{EXTRINSIC EFFECTS}
We now will focus on {\it extrinsic} broadening of photoemission
structures such as the Fermi edge. Within photoemission this up to
now meant just the finite instrumental resolution, but using the
picture that Joynt put forward in his recent article, we now also
have to consider losses suffered by the photoelectron {\it after}
it has left the solid and is on its way to the detector. These
losses are directly comparable to the loss spectrum of a
reflection EELS measurement. Joynt argues that these losses more
than anything will severely distort any sharp feature, such as the
Fermi edge, especially in the case of badly conducting solids and
anisotropic materials. He claims, depending on the properties of
the material under study, that this extrinsic distortion can be so
dramatic that instead of observing a sharp Fermi cutoff in the
spectrum, the observed spectral distribution will look like that
expected for a material with a so called pseudogap at E$_F$. If
correct this would make photoemission unsuitable for the study of
the one electron Green's function of badly conducting solids such
as many of the colossal magnetoresistance materials and the High
Tc's, among many others, are.\\ Joynt substantiates his statement
by deriving an expression for the energy loss probability of an
electron once the electron has emerged from the surface. He
calculates, in a classical picture, the average energy lost due to
the interaction of the electron's time dependent electric field
acting on the polarizable metal left behind. The response of the
metal is approximated by a Drudelike behaviour. He then
distributes this average energy lost over an energy loss spectrum
using a probability distribution as a weighting function. The
basic assumption is that the Born approximation is valid so that
each electron suffers at most a single scattering event. This
results in:
\begin{equation}
P(\omega) = \frac{2e^2CL(\omega)}{\pi\hbar v\omega^2}
\end{equation}
in which $P(\omega)$ stands for the probability that the electron
loses an energy $\omega$, $C$ is a constant for which Joynt gives
a value of $\approx 2.57$, and $L(\omega)$ represents the loss
function
\begin{equation}
L(\omega) = \frac{\omega}{4\pi} {\rm
Im}\left(\frac{-1}{1+\epsilon(\omega)}\right).
\end{equation}
We agree to a large extend with this derivation except perhaps for
the constant C on which we will focus our attention later on in
this paper. Let us for the moment assume that the Born
approximation and Joynt's derivation is correct. We see that the
material related properties enter the equation for the energy loss
distribution via the frequency dependent dielectric constant of
the material. Which is not unexpected since this describes the
response of the material to the time dependent field produced by
the outgoing electron. Let us, as Joynt did, take as an example
the dielectric function $\epsilon(\omega)$ given by the Drude
model:
\begin{equation}
\epsilon(\omega) = \frac{4\pi i}{\omega} \sigma(\omega)=\frac{4\pi
i \sigma_0}{\omega(1-i\omega\tau)}
\end{equation}
and plot $P(\omega)$ for different values of the resistivity
$\rho_0=1/\sigma_0$ and scattering time $\tau$ (Fig. 1) one sees
immediately that although the weight of $P(\omega)$ is distributed
differently in each curve, the integral $\int_0^\infty
P(\omega)d\omega$ reaches the same value in the end.
There is indeed a well known general sum rule\cite{Mahan}
connected to this formula which is independent of material
parameters:
\begin{equation}
\lim_{k\rightarrow 0}\frac{e^2 C}{2\pi^2\hbar
v}\int_0^\infty\frac{d\omega}{\omega}{\rm Im}\left(
\frac{-1}{1+\epsilon(\omega)}\right) = \frac{e^2 C}{2\pi^2\hbar
v}\frac{\pi}{2}\frac{1}{1+\epsilon_\infty}
\end{equation}
This is a general sum rule that holds for {\it any} model of
$\epsilon(\omega)$ provided that it's a causal function. It
depends only on  the velocity of the outgoing photoelectron which
is presumed constant in the process, consistent with the Born
approximation valid for weak scattering. So, for an electron with
a kinetic energy of 20 eV, as used in Joynt's examples, the
integral from zero to infinity gives $\approx 0.0843$ for the {\it
total} of the losses.\\ There is of course another sum rule which
conserves the integrated intensity of the spectrum. In this sum
rule we must include the finite probability $P_0$ that an electron
suffers no loss at all, so that
\begin{equation}
1 = P_0 + \int_0^\infty P(\omega) d\omega.
\end{equation}
Using the results above we find that $P_0\approx 0.957$, and
therefore the normal single particle DOS will dominate the
photoemission spectrum.\\ Up to now we've just been concentrating
on the free electron part of the response function, of course
there are more contributions (such as phonons, and interband
transitions) that are contained in the total dielectric function
of a real material. Therefore the calculations presented by us and
by Joynt, using only the Drude model, will always overestimate the
influence of the free electron contribution.\\ If we take a simple
example to illustrate this: our first sum rule pins down the total
amount of losses from zero to infinite frequency, this means that
other processes such as phonons, not contained in the Drude model
will just steal away some of the weight carried by the free
electron losses that we have considered so far. This implies that
in a good metal, where the phonon part will be nearly fully
screened by the free electron excitations, the (low energy) loss
spectrum is in first approximation indeed well described by the
Drude model. If the material is then changed into a bad conductor
the phonon part of the loss spectrum will gain more and more
strength in the low energy region of the spectrum at the expense
of the Drude part because of the sum rule. In both cases however,
P$_0$ will have a fixed value. Furthermore, any phonon
contribution will not influence the PES spectrum in a smooth way
on the low energy scale around the Fermi energy, but rather
produce a step in the convoluted photoemission-EELS spectrum,
since these phonon losses are peaked at the phonon frequency, and
are never overdamped. The same holds for a good metal where there
is a clearly defined loss peak at the plasma frequency.\\ Thus, we
can conclude that at 20 eV, $\int_0^\infty P(\omega)d\omega\approx
0.0843$ is an upper limit for the losses due to only the Drude
part. This very general result is in direct conflict with the
assumptions made by Joynt that $P_0$ could, within the
approximations made by him, be very small, as he hints at the fact
that processes other than the free electron losses will further
reduce $P_0$, and therefore he takes $P_0$ to be a fit parameter.
In our opinion it is not possible to make an independent choice
for the value of $P_0$ as Joynt did since it is in essence
determined by the sum rules and is independent of material
constants.\\ From an experimental point of view, we know from
reflection EELS experiments (see {\it e.g.}, Fig. 4) at incoming
energies of around 20 eV, that the elastic peak (which represents
$P_0$) is by no means close to zero for any material. Only for
very low incoming energies (below $\approx$ 10 eV) or when special
surface waveguidelike  conditions are met\cite{Pothuizen} can the
zero loss peak be strongly suppressed. Besides this, in reflection
EELS $P(\omega)$ is twice as strong as the loss probability in
this photoemission scenario, since the electron can lose energy
both on the incomming and on the outgoing trajectory. In fact, one
can use the same method as that used  in calculating the
reflection EELS loss probability (see Ibach \& Mills\cite{Mills})
for this photoemission problem. It is interesting to note that one
gets the same result except for a different numerical
factor.\cite{Mills2}. This  presumably stems from a difference in
Fourier transform convention and in Mills's case, avoiding
integrals such as equation (1) in Joynt's paper, which is not
readily solvable analytically. Using the prefactor obtained by the
procedure described by Mills the equation for the losses reads:
\begin{equation}
\label{Pmills}
 P(\omega) = \frac{e^2}{\hbar\omega v}{\rm
Im}\left(\frac{-1}{1+\epsilon(\omega)}\right)
\end{equation}
Which then for the sum rule means that the losses are in fact much
more severe than with Joynt's original prefactor: we now have
$\int_0^\infty P(\omega)d\omega \approx 0.65$ again at a kinetic
energy of the electron of 20 eV, leaving $P_0$ at 0.35. Even here,
the zero loss part will still be large enough to dominate the
Fermi cutoff, if one thinks in terms of the {\sl complete}
dielectric function being involved.\\ This result does imply
however that working in the Born approximation is no longer valid,
and a strong interaction picture containing also multiple losses
needs to be applied, which is less straightforward to derive for a
continuous spectrum of excitations. In the case of discrete,
welldefined plasmon losses due to core level excitations, the loss
spectrum has in fact been calculated by Langreth\cite{Langreth}
with the result that multiple plasmon losses are seen distributing
the energy loss over a wider energy range at the expense of the
low energy losses. On the basis of this calculation we argue that
multiple scattering corrections will not strongly influence the
sum rules but instead they will just redistribute the losses over
a wider energy range thereby reducing their influence in the low
energy loss region.\\ If we for the moment stick to the single
scattering scenario and take for the losses only the Drude
contribution, we can calculate the photoemission spectrum for the
same parameters as Joynt used, except now taking $P_0 = 0.35$.
This is depicted in Fig. 2, the lower panel also contains the
original calculation from Joynt, using $P_0 = 0.01$. From this we
see, that although the lineshape of the photoemission spectrum is
affected by the losses in the case of the bad metal (where the
Drude contribution is overestimated!), there is still finite
weight at the Fermi Energy and therefore the effect does not
create a clear pseudogap structure.
\section{EXPERIMENT}
As a last discussion point, we will present the case of
La$_{1.2}$Sr$_{1.8}$Mn$_2$O$_7$, a double layered colossal
magnetoresistance oxide with a ferromagnetic metal (at low
temperature) to paramagnetic insulator transition at 125 K grown
by the traveling solvent floating zone method\cite{Sury}. This
material is a good candidate for testing Joynt's assumption that
materials with high resistivity will be more prone to the
influence of losses on the Fermi region in photoemission spectra
than those with low resistivity, as the resistivity changes by
roughly two orders of magnitude from below to just above the phase
transition (see Fig.3)\cite{Sury}.
We performed both reflection EELS at 20.5 eV incoming electron
energy and angle integrated PES using a HeI source. For both
measurements the sample was cleaved {\sl in situ} at a temperature
of 60K, and at a base pressure of $8\times 10^{-11}$ in the case
of EELS and $4\times 10^{-11}$ in the case of PES. As these
samples deteriorate even at these pressures in a matter of hours,
we ensured that measurements were performed within 2 hours after
the cleave, before the peak at 9 eV binding energy in the PES
spectra started appearing, which is associated with a change in
oxygen stoichiometry at the surface\cite{Saitoh}.\\ The satellite
and background corrected regions around E$_F$ of the photoemission
spectra are shown in Fig. 5 for T=60K (solid), 140K (dashed) and
T=180K (dotted). The inset depicts the full spectra, taken at 60K
before and after the temperature cycle.\\ Since we performed EELS
at a finite incoming angle with respect to the surface normal
($\theta_{in} = 35\deg$) we have to apply a correction factor as
described by Ibach and Mills\cite{Mills} to extract the surface
loss function ${\rm Im}\left(-1/1+\epsilon(\omega)\right)$, and
then multiply this by $e^2/\hbar\omega v$ to get $P(\omega)$ as
described in equation (\ref{Pmills}) in order to use it to
simulate a PES spectrum. Therefore, in Fig. 4, top panel are
depict the EELS spectra as taken at 50K (solid) and 150K (dashed)
including the zero loss line, and in the lower panel $P(\omega)$
calculated from the data after subtraction of the zero loss line.

We then use this $P(\omega)$ to calculate its influence on a PES
spectrum assuming a constant density of states, and using various
values of $P_0$. This is depicted in the lower panel of Fig. 5. It
can be seen from this figure, that unless $P_0 = 0$ the picture of
Joynt doesn't reproduce the photoemission spectrum at all. For any
finite value of $P_0$ combined with a finite DOS at the Fermi
energy, one will get a finite Fermi cutoff.
To get a rough estimate of $P_0$ from experiment, we can integrate
the zero loss peak separately and compare it to half the integral
of the entire loss region up to 10 eV as measured in our EELS
spectra, provided of course that we keep in mind that we used our
detector in a mode which selects electrons within an opening angle
of $2^\circ$ around the specular reflection and is therefore not
fully angle integrated, which makes us underestimate the losses
relative to the zero loss probability. However, if we proceed in
this way, we get for both the 50 and 150K a ratio of
$P_0:P(\omega) = 0.82:0.18$ which shows at least that $P_0$ is not
close to zero, and neither does $P_0$ in our experiments depend on
temperature (or resistivity) of the sample. Our findings agree
with ARPES measurements by Dessau and Saitoh {\it et
al.}\cite{Dessau,Saitoh} in which they use Joynt's argument that
he expects the loss effect to be angle independent to show that
therefore the angle at which the smallest change is observed in
going from above to below Tc is indicative of the maximum
magnitude of the effect, and in their experiments turns out to be
negligible.
\section{CONCLUSIONS}
In conclusion, we have argued
that Joynt indeed raises an important question regarding the
influences of extrinsic losses on low energy photoelectrons, but
we disagree with the statement that the losses will take the upper
hand in determining the shape of the spectrum around the Fermi
energy as there is a sum rule that renders the zero loss
probability substantially finite. This at least holds down to
photon energies such as the often used HeI line (21.22 eV), but
may become a point of concern when really low photon energies are
used. Of course, since the losses are by no means a small
perturbation in this classical, single scattering approach a full
quantum mechanical treatment including multiple scattering is
called for. Furthermore, we are not able to reproduce Joynt's
formula exactly as far as the pre-factor is concerned, we believe
however that the treatment by Mills~\cite{Mills2} is
self-consistent and avoids integrals with questionable
convergence. We also have shown in the case of
$La_{1.2}Sr_{1.8}Mn_2O_7$ that we cannot reproduce the
photoemission spectrum using a finite density of states up to the
Fermi energy together with a finite value for $P_0$, so that we
must conclude that there is a true pseudogap in this material both
below and above the phase transition.\\
\begin{center}
{\bf ACKNOWLEDGEMENTS}
\end{center}
We would like to thank D. L. Mills for letting us use his
calculation of $P(\omega)$ for our simulations, and M. Mostovoy
and L. H. Tjeng for their valuable contributions along the way.
This research was supported by the Netherlands Foundation for
Fundamental Research on Matter (FOM) with financial support from
the Netherlands Organization for the Advancement of Pure Research
(NWO). The research of MAJ was supported through a grant from the
Oxsen Network.
\begin{figure}[h!]
\centerline{\epsfig{figure=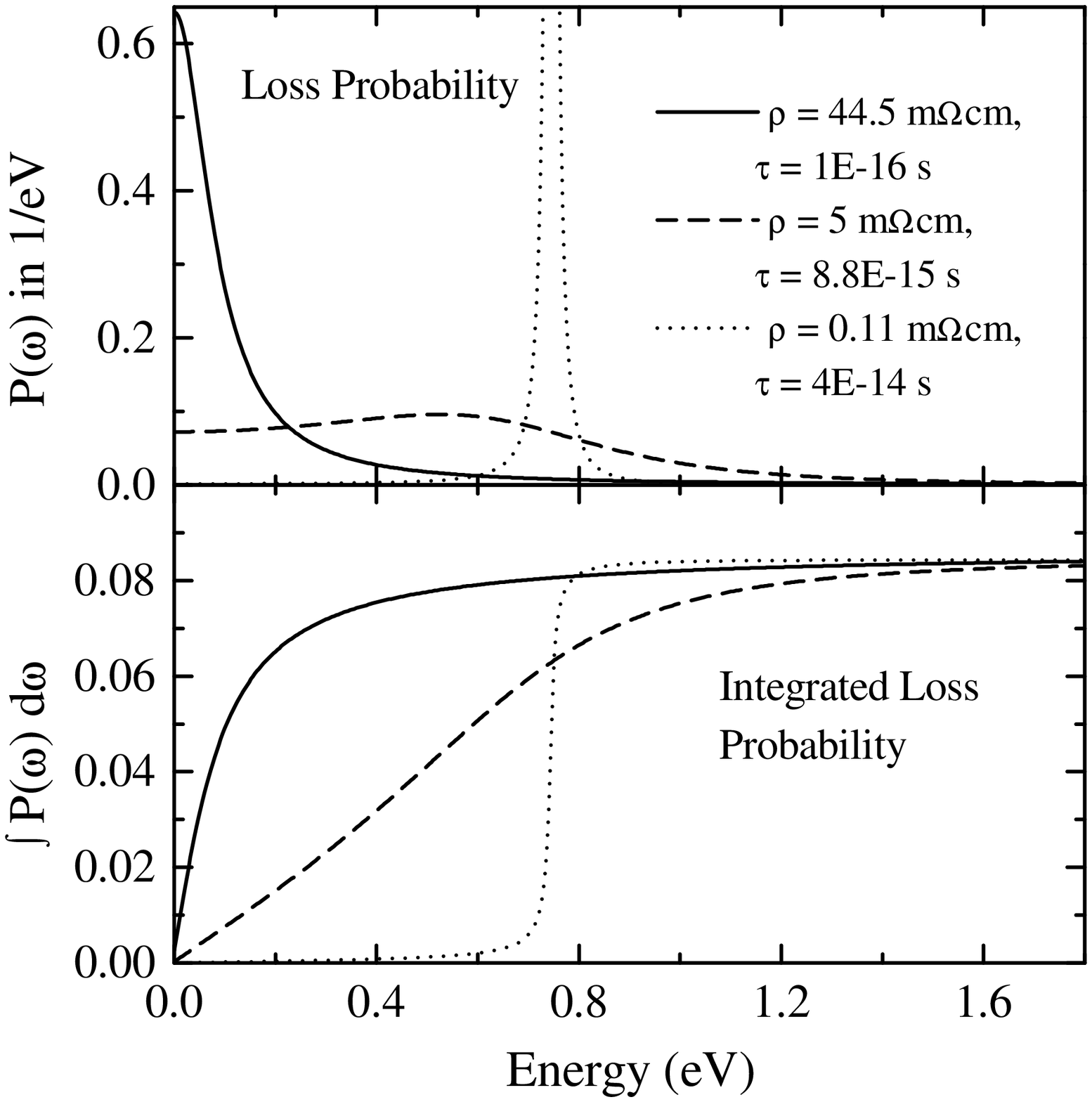,width=8.4cm,clip=}} \caption{Top
panel: The loss probability P($\omega$) calculated for different
values of resistivity $\rho$ and relaxation time $\tau$, keeping
the surface plasmon energy $\omega_{SP}^2 =
4\pi/(\rho\tau(1+\epsilon_{\infty}))$ fixed at 0.75 eV. Lower
panel: Demonstration of the sumrule for the loss probability
P($\omega$).} \label{joynt}
\end{figure}
\begin{figure}[h!]
\centerline{\epsfig{figure=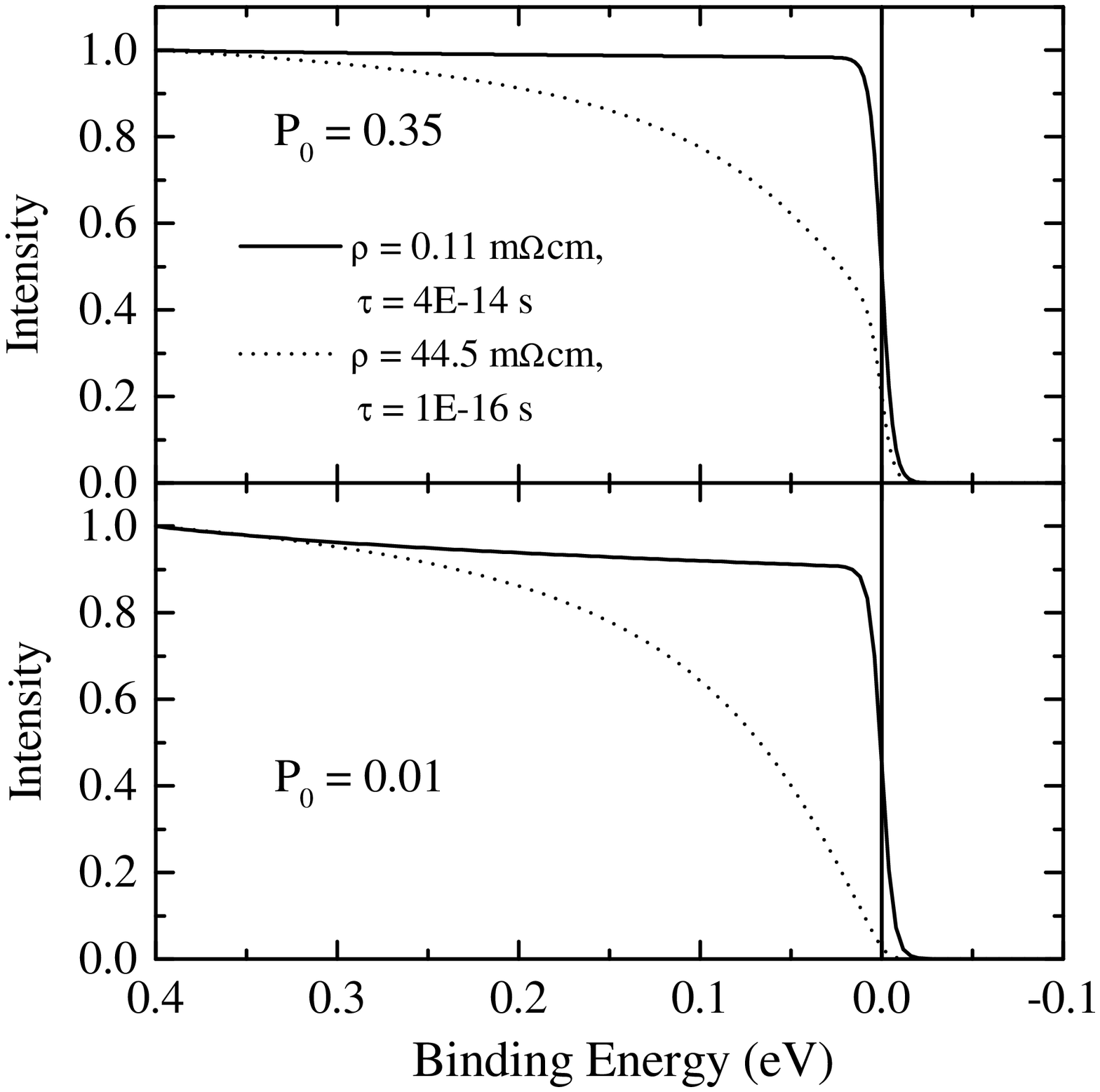,width=8.4cm,clip=}}
\caption{Comparison between Joynt's original calculation with $P_0
= 0.01$ and $\int P(\omega)d\omega = 0.0843$ (lower panel) and our
calculation  using Mills' prefactor for the loss probability
P($\omega$) and obeying the sum rule $P_0 = 1-\int
P(\omega)d\omega = 0.35$ (top panel) The parameters are specified
in the figure and T = 38K. Although the shape of the spectrum is
affected, there is no clear sign of a pseudogap for the bad metal
in our case.} \label{millsvsjoynt}
\end{figure}
\begin{figure}[h!]
\centerline{\epsfig{figure=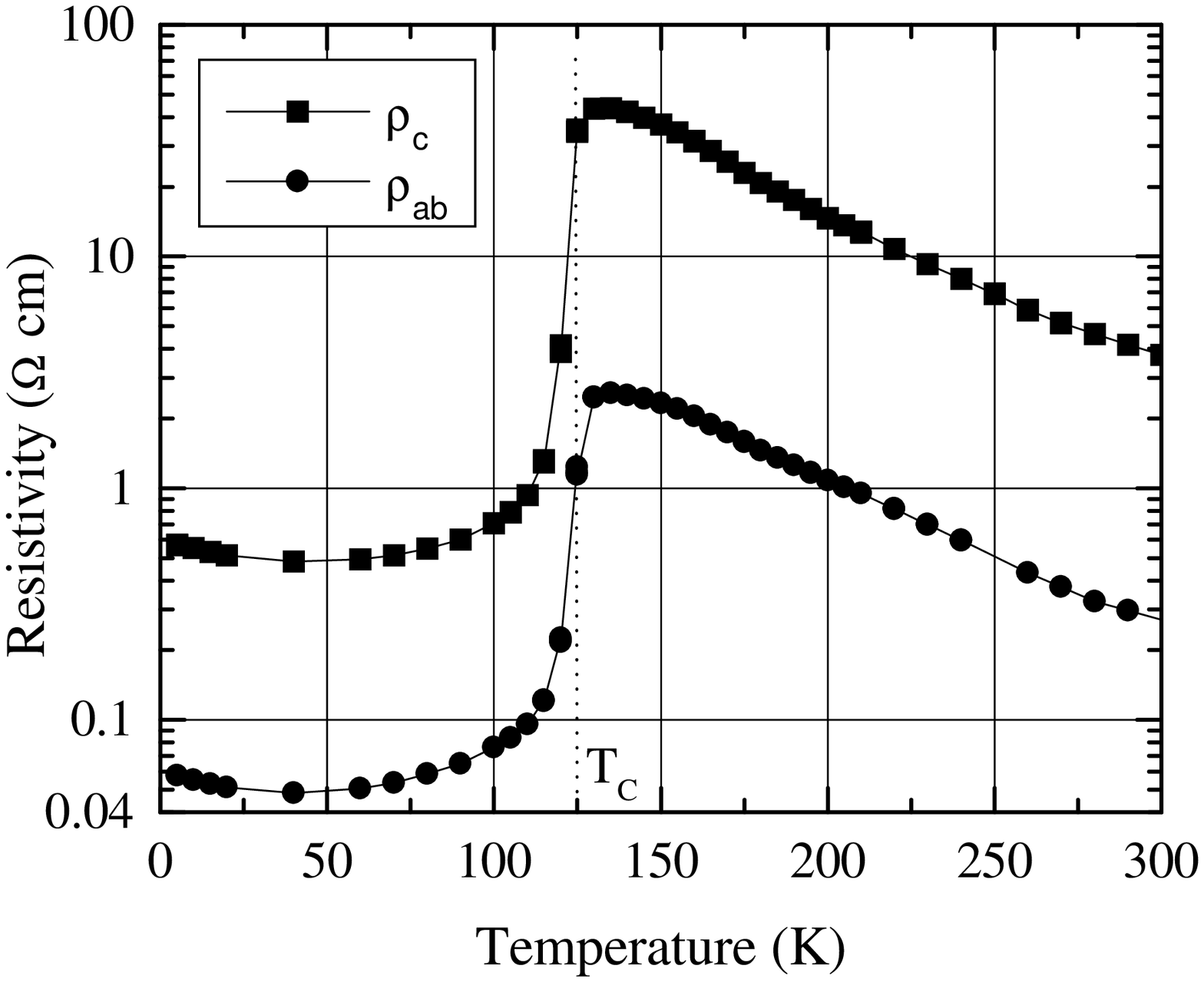,width=8.4cm,clip=}} \caption{The
resistivity of our $La_{1.2}Sr_{1.8}Mn_2O_7$ samples in the ab
plane (circles) and the c direction (squares) as a function of
temperature (from Ref. 9).} \label{lasrrho}
\end{figure}
\begin{figure}[h!]
\centerline{\epsfig{figure=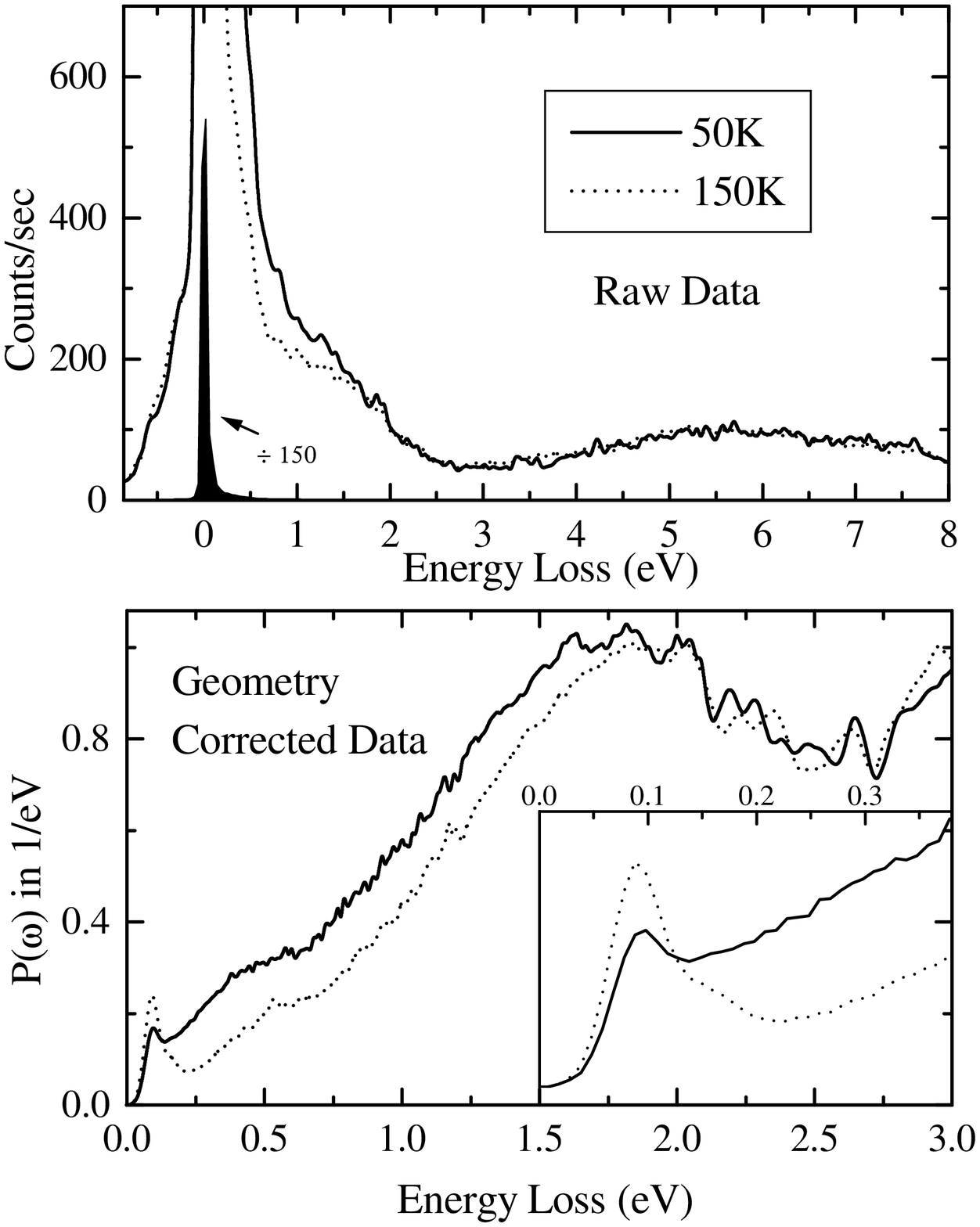,width=8.4cm,clip=}} \caption{Top
panel: Reflection EELS spectra, as taken, of
$La_{1.2}Sr_{1.8}Mn_2O_7$, at 50K (solid) and 150K (dotted).
Incoming/outgoing angle is $35^\circ$ with respect to the surface
normal. Incoming electron energy is 20.5 eV. Zero Loss FWHM $= 60$
meV. The filled black curve is the 50K spectrum divided by 150 to
show the relative weakness of the loss features with respect to
the zero loss electrons. Lower panel: Geometry corrected spectra,
according to Ref. 6 multiplied by $e^2/\hbar\omega v$ to get
P($\omega$). Inset: first 400 meV, showing the stronger presence
of a surface phonon in the insulating regime, relative to the
metallic phase.} \label{lasreels}
\end{figure}
\begin{figure}[h!]
\centerline{\epsfig{figure=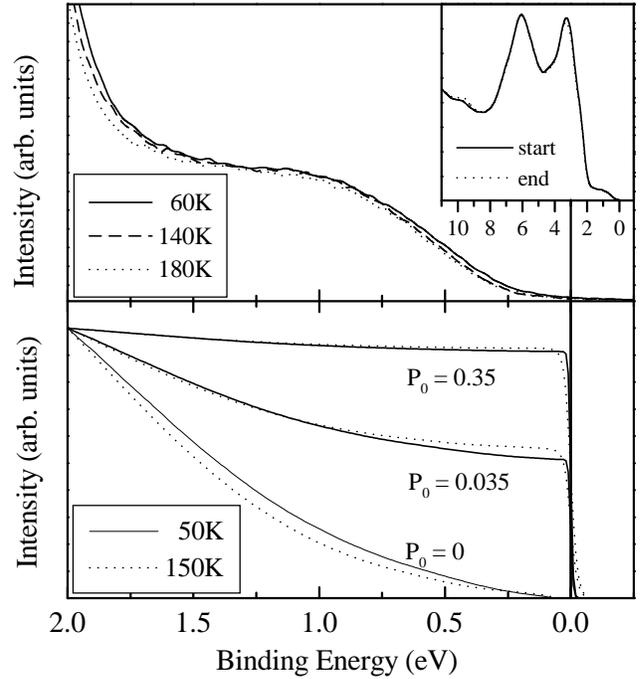,width=8.4cm,clip=}} \caption{Top
panel: PES spectra at different temperatures above and below the
phase transition in $La_{1.2}Sr_{1.8}Mn_2O_7$ close to the `Fermi
energy', the inset shows the spectra up to 12 eV binding energy,
before and after the temperature cycle ensuring the sample has not
degraded in the mean time. Lower panel: Calculated PES spectra
using the loss probability P($\omega$) constructed from the EELS
data of Fig. 4 and assuming a constant DOS, for various values of
the zero loss probability P$_0$ as indicated in the figure.}
\label{lasrpes}
\end{figure}
\end{document}